\documentclass[preprint]{aastex}
\usepackage{natbib}
\usepackage{graphicx}

\begin{document}

\title{Statistical Mechanics of the Cosmological
Many-body Problem and its Relation to Galaxy Clustering}

\author{William C. Saslaw}
\affil{Department of Astronomy, University of Virginia, USA and Institute of Astronomy, Cambridge, UK}

\author{Abel Yang}
\affil{Department of Astronomy, University of Virginia, USA}

\begin{abstract}
The cosmological many-body problem is effectively an infinite system of gravitationally interacting masses in an expanding universe. Despite the interactions' long-range nature, an analytical theory of statistical mechanics describes the spatial and velocity distribution functions which arise in the quasi-equilibrium conditions that apply to many cosmologies. Consequences of this theory agree well with the observed distribution of galaxies. Further consequences such as thermodynamics provide insights into the physical properties of this system, including its robustness to mergers, and its transition from a grand canonical ensemble to a collection of microcanonical ensembles with negative specific heat.
\end{abstract}

\section{Introduction}
\label{s1-intro}

\begin{quotation}
\center
\textit{E pluribus unus} \\
(One composed of many: Virgil, Moretum, 1, 104)\\
\ \\
\end{quotation}

Imagine an expanding universe filled with objects moving under their mutual gravitational forces. What decides the distribution of these objects in space, and their velocities, after the memory of their initial state has faded?  Richard Bentley, England's leading 17th century classicist essentially posed the spatial part of this question to his friend Isaac Newton. Their ensuing letters may be found today in the library of Trinity College, Cambridge, and are discussed in some detail elsewhere \citep{2000dggc.conf.....S}, along with the subject's subsequent history. Briefly, Newton surmised that if the universe were finite the objects would eventually all fall together into one large cluster. But if the universe were infinite, they ``could never convene into one mass; but some of it would convene into one mass and some into another, so as to make an infinite number of great masses, scattered at great distances from one to another throughout all that infinite space.'' And there the question rested for almost 300 years.

Today we restate this question by replacing ``finite'' and ``infinite'' by ``static'' and ``expanding'' and talking about galaxies instead of stars. The cosmological many-body problem appears to be a major component of the clustering of galaxies, although a detailed analysis of the astronomical observations also involves the roles of dark matter, galaxy formation and evolution, and perhaps dark energy.

Early investigations of the gravitational many-body problem used a fluid approach when \citet{1902RSPTA.199....1J} explored the stability of a self-gravitating gas. In the cosmological problem, the fundamental particles are usually galaxies. Jeans' results described gravitational collapse, and gave a timescale, proportional to $(G \rho)^{-1/2}$, for the collapse.

Jeans' solutions were developed for a static universe, since expansion had not been discovered. However, \citet{1930MNRAS..90..668E} showed that a static universe is unstable and a slight perturbation will cause it either to start expanding or collapsing. Hubble's \citeyearpar{1929PNAS...15..168H} discovery of expansion and the possible existence of dark energy \citep{1998AJ....116.1009R,1999ApJ...517..565P} provide a framework for the expanding universe and suggest that the universe will continue expanding so that structure formation and growth may eventually cease. This leads to interesting properties that we will explore below.

In addition to expansion, the other major discovery that simplifies the cosmological many-body problem is that the two-point correlation function of galaxies, $\xi_{2}(r)$, has a power law form $\xi_{2}(r) \propto r^{-\gamma}$ \citep{1969PASJ...21..221T}. This two-point galaxy correlation function is defined by the relation
\begin{equation} \label{1-eq1}
P(r|N_1=1)dr = 4\pi r^2 \overline{n}(1+\xi_2(r))dr
\end{equation}
for a statistically homogenous system of average number density $\overline{n}$. $P(r|N_1=1)$ is the conditional probability given a galaxy in an infinitesimal volume at an arbitrary coordinate origin, that there is another galaxy at a distance $r$. Conventionally, spherical volumes are used although this can easily be generalized to volumes of arbitrary shape which provide further information. Two-point correlation functions therefore represent an excess over a random Poisson probability. In our case this excess is caused by the galaxies' mutual attraction. For positive $\gamma$, $\xi_{2}(r)$ decreases significantly at large $r$. Observationally there is a correlation length, $R_1$, beyond which the correlation function decreases even faster than a power law, and may be neglected. Therefore over sufficiently large scales, the spatial distribution of galaxies is uncorrelated, and has a Poisson distribution, modified by its clustering on smaller scales. Modern surveys such as the 2DFGRS\citep{2003MNRAS.346...78H} have shown that 
\begin{equation} \label{1-eq1a}
\xi_{2}(r) \approx \left(\frac{r}{5.05h^{-1} \textrm{ Mpc}}\right)^{-1.67}
\end{equation}
with $R_{1} \approx 20 h^{-1}$ Mpc and $H_{0} = 100 h$ where $H_0$ is the Hubble constant. Amusingly, the exponent and amplitude in equation (\ref{1-eq1a}) are very close to those found in Totsuji and Kihara's analysis of the Lick survey, 40 years ago.

This result implies that on sufficiently large scales, the average density and the gravitational mean field is constant and isotropic. The resultant net force on a galaxy from the distant universe is negligible. On local scales, galaxies are subject to the long range effects of gravity from their neighbours within $R_1$.

For an expanding universe, \citet{1996ApJ...460...16S} showed that the expansion of the universe cancels the effect of the long range gravitational field for distances greater than $R_{1}$. This result is valid for Einstein-Friedmann models and for models that incorporate a cosmological constant, including $\Lambda$CDM cosmology. Because of expansion, the infinite range gravitational force has a finite effective range. Large-scale structures that form in the universe may be essentially stable as long as the universe is statistically homogenous at scales larger than $R_{1}$.

These results simplify the problem. Since the expansion of the universe effectively cancels the long range gravitational field, we can calculate the potential on a galaxy by integrating over a finite region of space instead of over the entire universe. Although the problem still involves an infinite volume, the gravitational field now has a finite range and we can consider a finite sub-volume caused by this cancellation.

Gravitational systems with more than two interacting particles are essentially unstable and not in equilibrium. While thermodynamics describes equilibrium systems, the cosmological case is characterised by quasi-equilibrium. This means that macroscopic quantities such as average temperature, pressure and density satisfy equilibrium relations whose variables change slowly compared with local relaxation timescales. For example, the average density of the universe(including dark matter) is approximately $\overline{\rho} = 3.5\times10^{10} m_{\odot}\textrm{Mpc}^{-3}$ \citep{2007ApJS..170..377S}, and an average galaxy has a mass of approximately $10^{11} m_{\odot}$. Hence the dynamical timescale of the universe is $\tau_{\textrm{universe}}\approx 25$ Gyr, and there are about $0.35$ galaxies per cubic megaparsec. The average peculiar velocity of a galaxy is $\sim 1000$ km s$^{-1}$, or 1 Mpc Gyr$^{-1}$. Therefore a cube with sides on the order of $R_{1}$ will have about 3000 galaxies, and the time for a galaxy to cross the cube is about 20 Gyr, or 1.5 times the age of the universe.

Local timescales are much shorter. A typical cluster similar to our local group of galaxies has a density(including dark matter) $\rho_{LG} \approx 1.5\times10^{12}m_{\odot} \textrm{Mpc}^{-3}$. This is $\sim 40$ times greater than $\overline{\rho}$, and hence the local group has a dynamical timescale $\tau_{LG} \approx 40^{-1/2} \tau_{\textrm{universe}} \approx 4$ Gyr. Rich clusters which are much denser than the local group will have correspondingly shorter dynamical timescales. Most clusters have an average diameter of $2\sim4$ Mpc, and hence the time taken to cross from one end to another is on the order of $3$ Gyr. These numbers mean that ``microscopic'' perturbations on local scales will relax significantly faster than the ``macroscopic'' scale of a cube of side $R_{1}$, so that the system generally evolves from one equilibrium state to another in quasi-equilibrium.

\subsection{The GQED}

The preceding considerations frame the problem. We have a gravitational field with an effective range $R_{1}$ in a system that is in quasi-equilibrium for which we need the counts-in-cells distribution $f(N)$, \textit{i.e.} the probability that a given cell located randomly in space contains $N$ galaxies. Thermodynamics and statistical mechanics provide two approaches to this problem. The thermodynamic solution was the first to be investigated \citep{1984ApJ...276...13S} and gives
\begin{equation} \label{1-eq2}
f(N;\overline{N},b) = \frac{\overline{N}(1-b)}{N!} \left[\overline{N}(1-b)+Nb\right]^{N-1}e^{-(\overline{N}(1-b)+Nb)}
\end{equation}
where $\overline{N}$ is the average number of galaxies in a cell, and $b$ is a clustering parameter equal to the ratio between the correlation potential energy and twice the kinetic energy:
\begin{equation} \label{1-eq3}
b = -\frac{W}{2K} = \frac{b_{0}\overline{n}T^{-3}}{1+b_{0}\overline{n}T^{-3}}
\end{equation}
with $0 \leq b \leq 1$. Here $T$ is the kinetic temperature of the system with kinetic energy $K = 3NT/2$ (taking the Boltzmann constant = 1), and $\overline{n}=\overline{N}/V$ is the average number of galaxies per unit volume where $V$ is the cell volume. The distribution given by equation (\ref{1-eq2}) describes the clustering of galaxies in quasi-equilibrium, and hence is known as the gravitational quasi-equilibrium distribution(GQED).

The galaxy spatial distribution function $f(N)$ is a simple but powerful statistic which characterises the locations of galaxies in space. It includes statistical information on voids and other underdense regions, on clusters of all shapes and sizes, on the probability of finding an arbitrary number of neighbours around randomly located positions, on counts of galaxies in cells of arbitrary shapes and sizes randomly located, and on galaxy correlation functions of all orders. These are just some of its representations \citep{2000dggc.conf.....S}. Moreover it is also closely related to the distribution function of the peculiar velocities of galaxies around the Hubble flow \citep{1990ApJ...365..419S,2004ApJ...608..636L}.

While the form of $b$ in the second equality of equation (\ref{1-eq3}) was originally taken as an ansatz by \citet{1984ApJ...276...13S}, a physical reason for the form was given by \citet{1996ApJ...460...16S} using constraints from thermodynamics and the boundary conditions of the problem. Writing the correlation potential energy in terms of the two-point correlation function gives
\begin{equation} \label{1-eq4}
b = -\frac{W}{2K} = \frac{2 \pi G m^{2}\overline{n}}{3T}\int_{V}\xi_{2}(r)r dr
\end{equation}
where $m$ is the mass of an individual galaxy.

The integral in equation (\ref{1-eq4}) indicates that $b$ depends on the shape and size of the cell, often taken to be spherical for simplicity. More generally, the shape of a cell affects the correlation potential energy so that\citep{2000dggc.conf.....S}
\begin{equation} \label{1-eq4a}
b = \frac{G m^{2}\overline{n}^{2}}{6 \overline{N}T}\int_{V}\int_{V}\frac{\xi_{2}(|\mathbf{r}_{1} - \mathbf{r}_{2}|)}{|\mathbf{r}_{1} - \mathbf{r}_{2}|} d\mathbf{r}_{1} d\mathbf{r}_{2}.
\end{equation}

\subsection{The Statistical Mechanical Approach}

The statistical mechanical approach to this problem was originated by \citet{2002ApJ...571..576A} who treated the cells as a grand canonical ensemble with the universe as an energy and particle reservoir and the galaxies as particles in the ensemble. The statistical mechanical results are essentially the same as those in the thermodynamic analysis of point particles, but here we use their generalization to extended particles. In this section we consider the simplest case with the following assumptions:
\begin{itemize}
\item Galaxies behave like point masses and the two-galaxy interaction potential is $\phi(\mathbf{r}) = -G m^{2}/|\mathbf{r}|$. This will be generalised to include a softening parameter, $\epsilon$, at small scales in equation (\ref{2-eq1}). But since the lower limits of the integrals in equations (\ref{1-eq5})-(\ref{1-eq6a}) converge uniformly in the limit $\epsilon = 0$, we first use this limit to give the simple formulae in equations (\ref{1-eq7})-(\ref{1-eq13}) of the original thermodynamic results.
\item Only 2-body interactions are considered as the dominant type of interactions.
\item $\phi(\mathbf{r})T^{-1}$ is small.
\end{itemize}
All three assumptions can be relaxed and we will discuss the resulting modifications in section \ref{s2-mod}.

In order to compute the partition function, we note that since expansion cancels the long range mean-field force the integral over physical space has a finite cutoff and the integral
\begin{equation} \label{1-eq5}
\int_{0}^{R_{1}} \exp\left[-\phi(\mathbf{r})T^{-1}\right] d^{3}\mathbf{r} = \int_{0}^{R_{1}} \exp\left[\frac{G m^{2}}{|\mathbf{r}| T} \right] d^{3}\mathbf{r}
\end{equation}
for a galaxy mass potential has no singularities if the third assumption above holds.

In order to compute the partition function, we start with the normalised phase space integral
\begin{eqnarray} \label{1-eq6}
Z_{N}(T,V) &=& \frac{1}{\Lambda^{3N}N!}\int\exp\left[-\left(\sum_{i=1}^{N}\frac{\mathbf{p}_{i}^{2}}{2m}+\phi(\mathbf{r}_{1},\ldots,\mathbf{r}_{N})\right)T^{-1}\right] d^{3N}\mathbf{p} d^{3N} \mathbf{r} \nonumber \\
 &=& \frac{1}{N!}\left(\frac{2 \pi m T}{\Lambda^{2}}\right)^{3N/2} Q_{N}(T,V)
\end{eqnarray}
with $\Lambda$ a normalisation constant and $V$ the volume of the ensemble. From the observation that most gravitational interactions are dominated by two-body interactions, and by taking the first order expansion of the exponential, the potential energy term of the configuration integral $Q_{N}(T,V)$ becomes
\begin{eqnarray} \label{1-eq6a}
Q_{N}(T,V)&= & \int\exp\left[-\left(\phi(\mathbf{r}_{1},\ldots,\mathbf{r}_{N})\right)T^{-1}\right] d^{3N} \mathbf{r} \nonumber \\
&\approx& \int\prod_{1 \leq i < j \leq N}\exp\left[-\phi(\mathbf{r}_{ij})T^{-1}\right] d^{3N} \mathbf{r} \nonumber \\
&\approx& \int\prod_{1 \leq i < j \leq N}\left[1+\frac{G m^{2}}{T|\mathbf{r}_{ij}|}\right] d^{3N} \mathbf{r},
\end{eqnarray}
assuming that galaxies at large distances behave like point masses(although their finite extent is important in close interactions and mergers), that 2-body interactions dominate, and that $\phi(\mathbf{r}_{ij})T^{-1}$ is small.
This gives the partition function for a canonical ensemble, the Helmholtz free energy, and the corresponding thermodynamic variables for entropy, pressure, internal energy and chemical potential:
\begin{equation} \label{1-eq7}
Z_{N}(T,V) = \frac{1}{N!}\left(\frac{2 \pi m T}{\Lambda^{2}}\right)^{3N/2} V^{N}\left(1-b\right)^{1-N}
\end{equation}
\begin{equation} \label{1-eq7f}
F = -T \ln Z_{N}(T,V) = NT \ln\left(\frac{N}{VT^{3/2}}\right)+NT\ln\left(
1-b\right)-NT-\frac{3}{3}NT\ln\left(\frac{2\pi m}{\Lambda^{2}}\right)
\end{equation}
\begin{equation} \label{1-eq7a}
S = -\left(\frac{\partial F}{\partial T}\right)_{V,N} = -N \ln\left(\frac{N}{VT^{3/2}}\right) - N\ln(1-b)-3Nb+\frac{5}{2}N+\frac{3}{2}N\ln\left(\frac{2 \pi m}{\Lambda^{2}}\right)
\end{equation}
\begin{equation} \label{1-eq7b}
P = -\left(\frac{\partial F}{\partial V}\right)_{T,N} = \frac{NT}{V}(1-b)
\end{equation}
\begin{equation} \label{1-eq7c}
U = F + TS = \frac{3}{2}NT(1-2b)
\end{equation}
\begin{equation} \label{1-eq7d}
\mu = \left(\frac{\partial F}{\partial N}\right)_{T,V} = T \ln\left(\frac{N}{VT^{3/2}}\right) + T\ln(1-b) - Tb -\frac{3}{2}T\ln\left(\frac{2 \pi m}{\Lambda^{2}}\right)
\end{equation}

In these derivations the functional form of $b$ occurs naturally as
\begin{equation} \label{1-eq8}
b = \frac{(3/2)G^{3}m^{6}\overline{n}T^{-3}}{1+(3/2)G^{3}m^{6}\overline{n}T^{-3}}.
\end{equation}
Comparing the coefficients of equations (\ref{1-eq8}) and (\ref{1-eq3}), we see that
\begin{equation} \label{1-eq9}
b_{0} = (3/2)G^{3}m^{6}
\end{equation}
which quantitatively relates the correlation potential energy to the mass of an individual galaxy and confirms the original ansatz in equation (\ref{1-eq3}).

\begin{figure}[tbp]
\begin{center}
\includegraphics[width=\textwidth]{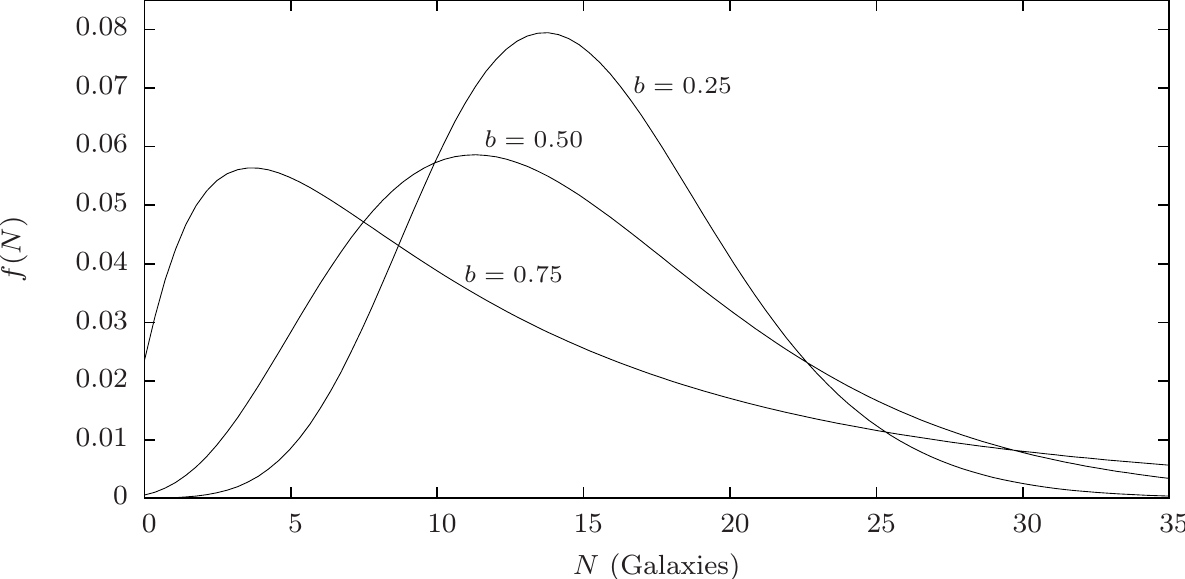}
\caption{The GQED from equation (\ref{1-eq2}) for $\overline{N}=15$}
\label{1-fig1}
\end{center}
\end{figure}

The distribution function is obtained by summing over all energy states for given $N$:
\begin{equation} \label{1-eq10}
f(N) = \frac{e^{N\mu/T}Z_{N}(T,V)}{Z_{G}(T,V)}.
\end{equation}
By using equations (\ref{1-eq7})-(\ref{1-eq7d}) and the relation between the grand canonical partition function $Z_{G}$ and the pressure equation of state, we then again obtain the distribution function given by equation (\ref{1-eq2}). Figure \ref{1-fig1} illustrates the counts in cells distribution for different values of $b$. For $b=0$, there is no interaction and equation (\ref{1-eq2}) becomes a Poisson distribution.

\subsection{The Peculiar Velocity Distribution}

The GQED also contains a consistent distribution function for the peculiar velocities of galaxies. Since the partition function separates into kinetic and potential energy factors, the phase space distribution is separable such that \citep{1990ApJ...365..419S}
\begin{equation} \label{1-eq11a}
f(N,v) \to f(N)h(v).
\end{equation}

To relate the counts-in-cells distribution to the peculiar velocity distribution, we need a relation between the number density $N$ and the velocity $v$. This arises from the proportionality between the fluctuations in potential energy (due to correlations) over a given volume and the local kinetic energy fluctuations:
\begin{equation} \label{1-eq11b}
G m N \left\langle \frac{1}{r}\right\rangle = \alpha v^{2}.
\end{equation}
Here $\alpha$ is a constant of proportionality such that $\alpha = \left\langle 1/r \right\rangle \left\langle v^{2}\right\rangle^{-1}$, $v$ is the peculiar velocity of a galaxy, and $r$ is the separation between two galaxies \citep{2004ApJ...608..636L}.

\begin{figure}[tbp]
\begin{center}
\includegraphics[width=\textwidth]{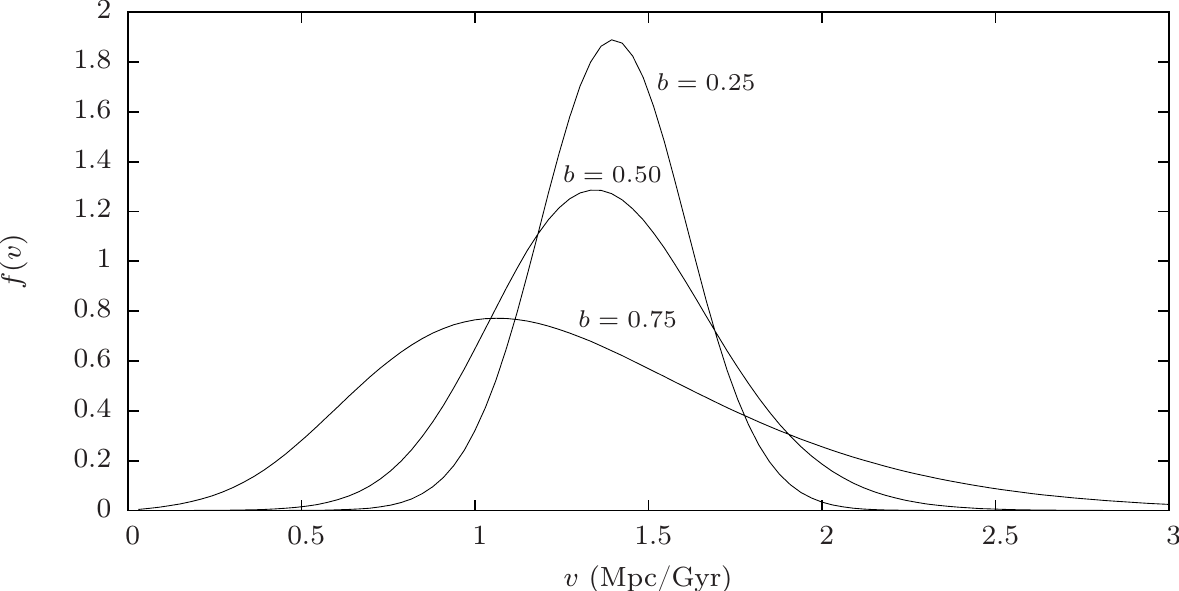}
\caption{The peculiar velocity distribution from equation (\ref{1-eq12}) for $\alpha=10$ and $\beta=2.0$ with units of velocity in Mpc Gyr$^{-1}$}
\label{1-fig2}
\end{center}
\end{figure}

With the relation between $N$ and $v$ in equation (\ref{1-eq11b}), the distribution in equation (\ref{1-eq2}) can be transformed to a distribution in $v$ \citep{1990ApJ...365..419S}:
\begin{equation} \label{1-eq12}
f(v) dv = \frac{2 \alpha^{2} \beta (1-b)}{\Gamma\left(\alpha v^{2}+1\right)}\left[\alpha\beta(1-b)+ \alpha b v^{2}\right]^{\alpha v^{2}-1} e^{-\left(\alpha\beta(1-b) + \alpha b v^{2}\right)} v dv
\end{equation}
where $\beta \equiv \left\langle v^{2}\right\rangle$, $\Gamma$ is the standard gamma function, and averages are taken over the grand canonical ensemble. Figure \ref{1-fig2} illustrates the peculiar velocity distribution for different values of $b$.

Since the proper motions of galaxies are too small to be observed, we can only measure the radial component of their peculiar velocity along our line of sight. We can generally write the velocity as a component parallel to our line of sight and a component perpendicular to our line of sight such that $v^{2} = v_{\parallel}^{2}+v_{\perp}^{2}$. Then to obtain a form of the velocity distribution directly comparable with observations, we integrate over all perpendicular velocities to get the radial velocity distribution function\citep{1992ApJ...386....9I}
\begin{eqnarray} \label{1-eq13}
f(v_{\parallel}) &=& \alpha^{2}\beta(1-b)e^{-\alpha \beta(1-b)}\int_{0}^{\infty}\frac{v_{\perp}}{\sqrt{v_{\parallel}^{2}+v_{\perp}^{2}}} \nonumber \\
 & &\times \frac{\left[\alpha\beta(1-b)+\alpha b \left(v_{\parallel}^{2}+v_{\perp}^{2}\right)\right]^{\alpha\left(v_{\parallel}^{2}+v_{\perp}^{2}\right)+1}}{\Gamma\left[\alpha\left(v_{\parallel}^{2}+v_{\perp}^{2}\right)+1\right]} e^{-\alpha b \left(v_{\parallel}^{2}+v_{\perp}^{2}\right)}dv_{\perp}.
\end{eqnarray}

Figures \ref{4-fig2} and \ref{4-fig4} below compare this with simulations and observations.

\section{Modifications and the General Form of the GQED}
\label{s2-mod}
Various modifications and extensions exist for the GQED. The simplest case involves a modification of the potential of a galaxy so that instead of treating galaxies as point masses, galaxies are treated as extended objects with a potential having the form \citep{2002ApJ...571..576A}
\begin{equation} \label{2-eq1}
\phi = -\frac{G m^{2}}{\left(r^{2}+\epsilon^{2}\right)^{1/2}}
\end{equation}
where $\epsilon$ is a softening parameter related to the radius of a galaxy. Such a modified potential is commonly used in $N$-body simulations to model extended galaxies possibly with dark matter halos, and to avoid the singularity in the point mass potential when the separation between galaxies is small. The result of such a softening parameter is a modification $b$ to $b_{\epsilon}$ such that
\begin{equation} \label{2-eq2}
b_{\epsilon} = \frac{(3/2)G^{3}m^{6}\zeta\left(\epsilon/R_{1}\right)\overline{n}T^{-3}}{1+(3/2)G^{3}m^{6}\zeta\left(\epsilon/R_{1}\right)\overline{n}T^{-3}}
\end{equation}
where $\zeta\left(\epsilon/R_{1}\right)$ is a term that depends on the interaction potential between a pair of galaxies. For $x \equiv \epsilon/R_{1}$
\begin{equation} \label{2-eq2a}
\zeta\left(x\right) = \sqrt{1+x^{2}} + x^{2}\ln\frac{x}{1+\sqrt{1+x^{2}}}.
\end{equation}

\begin{figure}[tbp]
\begin{center}
\includegraphics[width=\textwidth]{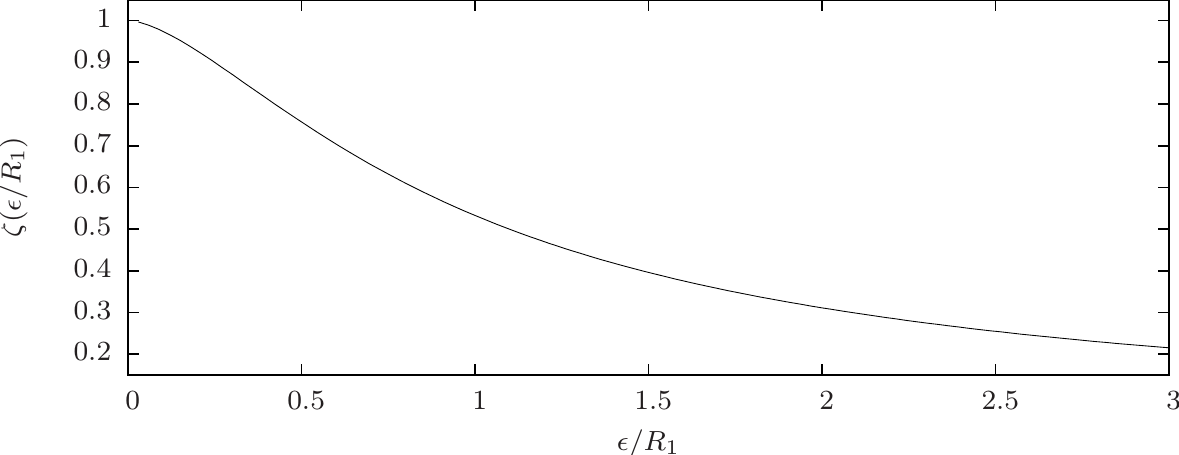}
\caption{$\zeta(\epsilon/R_{1})$ for the softened point mass potential given by equation (\ref{2-eq1}).}
\label{2-fig1}
\end{center}
\end{figure}

In the case of a point mass, $\epsilon\to 0$ and $\zeta\left(\epsilon/R_{1}\right) \to 1$ so the partition function and its consequent thermodynamic properties converge to the solution for a point mass potential. Figure \ref{2-fig1} illustrates $\zeta(\epsilon/R_{1})$.

To generalise this result, we can modify the point mass potential so that
\begin{equation} \label{2-eq3}
\phi = -\frac{G m^{2}}{r} \kappa(r)
\end{equation}
where $\kappa(r)$ is a dimensionless modification factor. It can then be shown that $\zeta(x)$ depends on the form of $\kappa(r)$ such that $\zeta(x) \to 1$ as $\kappa(r) \to 1$. This modification of the potential will only affect $b_{0}$. Therefore many other forms of the potential, such as multipole moments, are possible.

\subsection{Triplet Interaction}

\citet{2006ApJ...645..940A} investigated the contribution of the irreducible triplet interactions, which are the cases where three-body interactions occur very close together. In such a case, the modified partition function for a canonical ensemble of $N \geq 3$ galaxies is
\begin{equation} \label{2-eq5}
Z_{N}(T,V) = \frac{V^{N}}{N!}\left(\frac{2 \pi m T}{\Lambda^{2}}\right)^{3N/2}\left[\left(\frac{1}{1-b}\right)^{N-1} + \frac{(N-1)(N-2)}{2}\frac{4}{9}\left(\frac{b}{1-b}\right)^{3}\right]
\end{equation}
where we note that at large values of $N$, the $(b/(1-b))^{3}$ term becomes very small compared to the $(1/(1-b))^{N-1}$ term. This indicates that the contribution from irreducible triplets is negligible compared to pairwise interactions for large $N$.

From the partition function, the resulting distribution function taking into account irreducible triplets is
\begin{equation} \label{2-eq6}
f(N;\overline{N},b) = \frac{\overline{N}(1-b) \left[\overline{N}(1-b)+Nb\right]^{N-1}+N^{3}\overline{N}^{N-3}L(N)}{N!(1+L(N))} e^{-(\overline{N}(1-b_{t})+Nb_{t})}
\end{equation}
where we write  $a_{N} = 2(N-1)(N-2)/9$, and define $L(N)$ and $b_{t}$ as follows:
\begin{equation} \label{2-eq6a}
L(N) = \left\{\begin{array}{ll}a_{N} b^3 (1-b)^{N-3}, & \textrm{for }N \geq 3 \\0 & \textrm{for }N < 3\end{array}\right.
\end{equation}
\begin{equation} \label{2-eq6b}
b_{t} = \left\{\begin{array}{ll}b\left[\frac{N+3a_{N}b^{2}(1-b)^{N-3}}{N+Na_{N}b^{3}(1-b)^{N-3}}\right], & \textrm{for }N \geq 3 \\b & \textrm{for }N < 3\end{array}\right.
\end{equation}

\begin{figure}[tbp]
\begin{center}
\includegraphics[width=\textwidth]{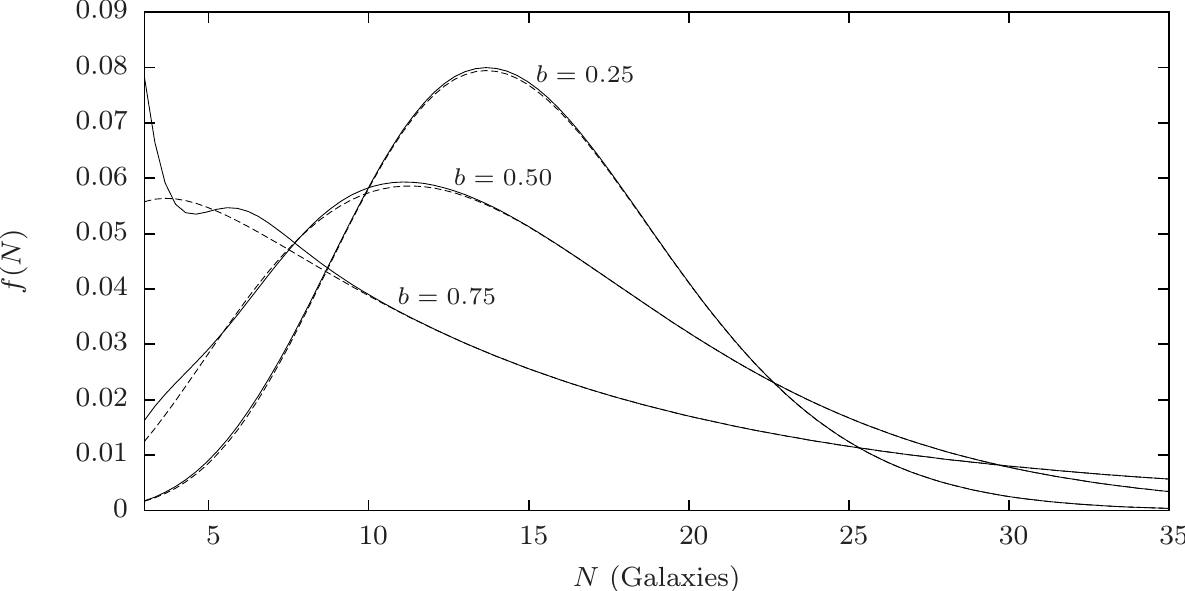}
\caption{The GQED including irreducible triplets (equation \ref{2-eq6}) for $\overline{N}=15$(solid line). The dashed line, for comparison, is the GQED (equation \ref{1-eq2}), without including irreducible triplets.}
\label{2-fig2}
\end{center}
\end{figure}

Figure \ref{2-fig2} compares the contribution of triplet interactions to the original GQED for different values of $b$.

\subsection{Two Species}

In order to help understand effects of a distribution of galaxy masses, \citet{2006IJMPD..15.1267A} derived the distribution function for a system consisting of two different galaxy masses and pointed out how this can be further generalised to a distribution of masses. We let the total number of galaxies in a cell be $N=N_{1}+N_{2}$ of which $N_{1}$ and $N_{2}$ have individual masses $m_{1}$ and $m_{2}$.

By considering interactions between the different species, the partition function for a canonical ensemble of $N$ galaxies becomes
\begin{equation} \label{2-eq8}
Z_{N}(T,V) = \frac{V^{N}}{\Lambda^{3N}N!}\left(2 \pi m_{1} T\right)^{3N_{1}/2}\left(2 \pi m_{2} T\right)^{3N_{2}/2}\left[1+b_{0}\overline{n}T^{-3}\right]^{N_{1}-1}\left[1+b_{12}\overline{n}T^{-3}\right]^{N_{2}}
\end{equation}
where for the case of a Newtonian point mass potential, $b_{0} = (3/2)(G m_{1}^{2})^{3}$, and $b_{12} = (3/2)(G m_{1}m_{2})^{3}$.

The distribution function is then
\begin{eqnarray} \label{2-eq9}
f(N;\overline{N},b) &=& \frac{\overline{N}(1-b)}{N!} \left[\overline{N}(1-b)+Nb\right]^{N_{1}-1} \nonumber \\
 & & \times \left[\frac{\overline{N}(1-b)+(m_{2}/m_{1})^{3}Nb}{1-b+(m_{2}/m_{1})^{3}b}\right]^{N_{2}} e^{-(\overline{N}(1-B)+NB)}
\end{eqnarray}
where $b$ is given as always by equations (\ref{1-eq3}) and (\ref{1-eq4}) and $B$ is given by
\begin{equation} \label{2-eq9a}
B = \frac{b}{(1+N_{2}/N_{1})}\left[1+\frac{(m_{2}/m_{1})^{3}(N_{2}/N_{1})}{1-b+(m_{2}/m_{1})^{3}b}\right].
\end{equation}

\begin{figure}[tbp]
\begin{center}
\includegraphics[width=\textwidth]{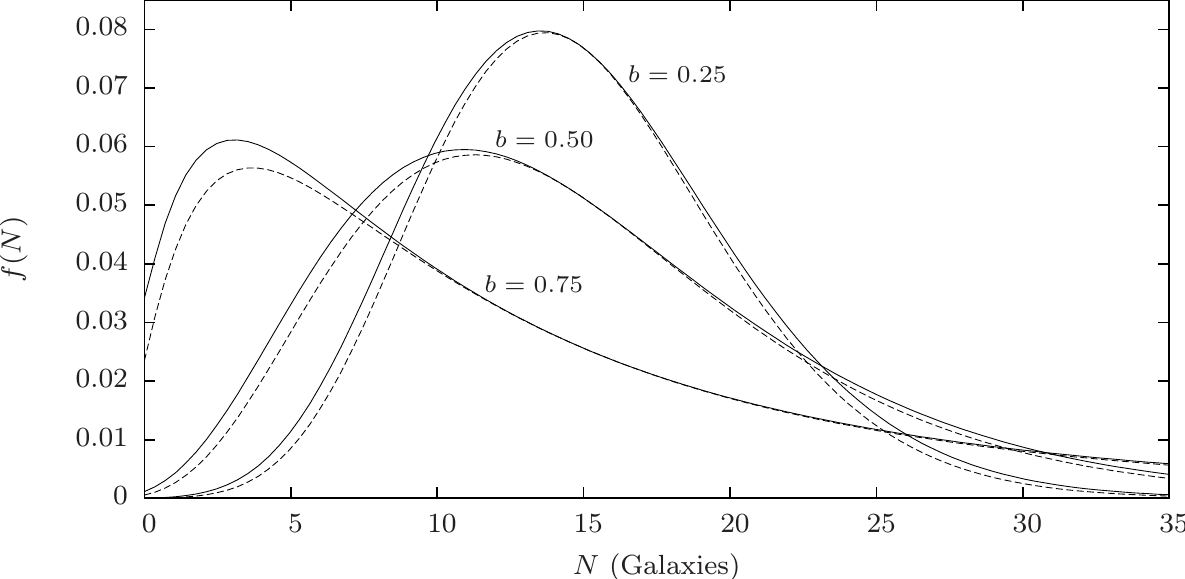}
\caption{The two species GQED from equation (\ref{2-eq9}) for $\overline{N}=15$(solid line) with a number ratio of $N_{2}/N = 0.1$ and mass ratio of $m_{2}/m_{1}=10$. The dashed line, for comparison, is the GQED(equation \ref{1-eq2}) with mass $m_1$ and $\overline{N}=15$.}
\label{2-fig3}
\end{center}
\end{figure}

We compare the 2-species distribution function with the original GQED in figure \ref{2-fig3}.

\subsection{Higher order expansions of $\exp(-\phi(r))/T$}
A higher order expansion of the exponential in equation (\ref{1-eq6a}) is considered by \citet{SA2009}. Again writing $b_{0} = (3/2)(G m^{2})^{3}$, the canonical partition function is now
\begin{equation} \label{2-eq11}
Z_{N}(T,V) = \frac{V^{N}}{N!}\left(\frac{2 \pi m T}{\Lambda^{2}}\right)^{3N/2}\left[1+b_{0} \overline{n}T^{-3}\zeta_{1}+\left(b_{0} \overline{n}T^{-3}\right)^{2}\zeta_{2}\right].
\end{equation}
where $\zeta_{1}$ and $\zeta_{2}$ are factors that arise from the second order expansion. In the case of point masses, $\zeta_{1}=1$ and $\zeta_{2}=2/3$.

A quantity $b_{\star}$ representing the modification of $b$ by the second order expansion can be defined as
\begin{equation} \label{2-eq12}
b_{\star} =
\frac{b_{0} \overline{n}T^{-3}\zeta_{1}+2\left(b_{0} \overline{n}T^{-3}\right)^{2}\zeta_{2}}
   {1+b_{0} \overline{n}T^{-3}\zeta_{1}+ \left(b_{0} \overline{n}T^{-3}\right)^{2}\zeta_{2}}
= \frac{b(1-b)\zeta_{1}+2b^{2}\zeta_{2}}{(1-b)^{2}+b(1-b)\zeta_{1}+b^{2}\zeta_{2}}
\end{equation}

The distribution function in this case is
\begin{equation} \label{2-eq13}
f(N;\overline{N},b) = \frac{\overline{N}(1-b)}{N!}\left[\frac{\overline{N}(1-b)+Nb\zeta_{1}+\frac{N^{2}b^{2}}{\overline{N}(1-b)}\zeta_{2}}{(1-b)+b\zeta_{1}+\frac{b^{2}}{(1-b)}\zeta_{2}}\right]^{N-1} \frac{e^{-(\overline{N}(1-b_{\star}) + N b_{\star})}}{(1-b)+b\zeta_{1}+\frac{b^{2}}{(1-b)}\zeta_{2}}.
\end{equation}

Normally the effects of the high order terms in this expansion are small, thereby confirming the essential form of $f(N)$ in equation (\ref{1-eq2}) and its consequences.

\section{Properties of the GQED}
\label{s3-prop}
The GQED has a number of properties that are discussed in detail in \citet{2000dggc.conf.....S}. We summarise some of them here.

\subsection{Correlation Functions}
For the simplest case, the two-point correlation function enters $f(N)$ through (\ref{1-eq4a}) since $b$ depends on the volume integral of $\xi_{2}$. However, in general the form of the GQED also contains information about the volume integrals of all the correlation functions where the average volume integral is defined as
\begin{equation} \label{3-eq4}
\overline{\xi}_N = \frac{1}{V^N}\int_V \xi_N(\mathbf{r}_1,\dots,\mathbf{r}_N) d^3\mathbf{r}_1\dots\mathbf{r}_N.
\end{equation}
For the general $N$-point correlation function, \citet{1989ApJ...343..107Z} derived a relation between $\overline{\xi}_N$, $\overline{N}$ and $b$. The general form for $\xi_N$ is
\begin{equation} \label{3-eq5}
\overline{\xi}_N = \frac{1-b}{b \overline{N}^{N-1}}\sum_{M=N}^{\infty} \frac{M^{M-1}}{(M-N)!}b^M e^{-Mb}.
\end{equation}
For $N=1$, 2 and 3 this gives
\begin{equation} \label{3-eq5a}
\overline{\xi}_1 = 1
\end{equation}
\begin{equation} \label{3-eq5b}
\overline{\xi}_2 = \frac{b}{(1-b)^2}\frac{2-b}{\overline{N}}
\end{equation}
\begin{equation} \label{3-eq5c}
\overline{\xi}_3 = \frac{b^2}{(1-b)^4}\frac{9-8b+2b^2}{\overline{N}^2}.
\end{equation}
Average moments of other quantities, including the thermodynamic variables, have also been derived.

\subsection{Specific Heat}
\begin{figure}[tbp]
\begin{center}
\includegraphics[width=\textwidth]{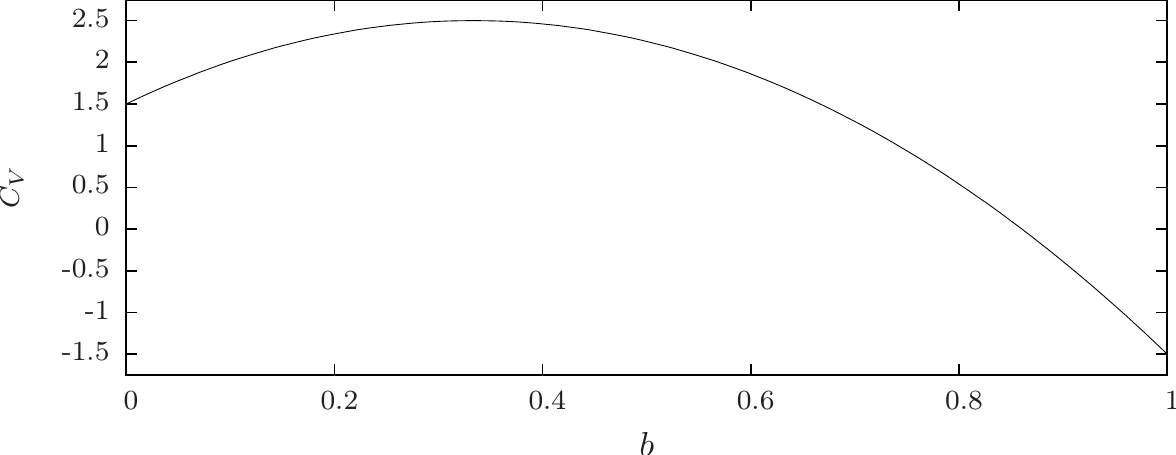}
\caption{The specific heat per galaxy at constant volume for varying $b$ from equation (\ref{3-eq1}).}
\label{3-fig1}
\end{center}
\end{figure}

From the internal energy given by equation (\ref{1-eq7c}), using equation (\ref{1-eq3}) the specific heat per galaxy at constant volume is
\begin{equation} \label{3-eq1}
C_V = \left.\frac{1}{N}  \frac{\partial U}{\partial T} \right|_{V,N} = \frac{3}{2}(1+4b-6b^2)
\end{equation}
which at $b=0$, is $3/2$ representing the monatomic ideal gas. As galaxies cluster, binaries start to dominate and $C_V$ reaches its maximum of $5/2$ at $b = 1/3$. At a critical value of $b_{crit}$ such that
\begin{equation} \label{3-eq2}
b_{crit} = \frac{2+\sqrt{10}}{6} = 0.8604
\end{equation}
$C_V = 0$. As $b$ increases beyond $b_{crit}$, $C_{V}$ decreases until $C_V$ reaches $-3/2$ at $b=1$ representing the specific heat of a fully virialized cluster. The transition from positive to negative specific heat, illustrated in figure \ref{3-fig1}, occurs at a rather high value of $b$, and does not involve a discontinuity, unlike some laboratory systems.

While negative specific heat is frequently encountered in the gravitational context, the system we are discussing is nominally a grand canonical ensemble. Grand canonical ensembles can only have positive specific heat\citep[equation 141.12]{TOL1938}. However for a microcanonical ensemble, the specific heat can be negative\citep{1970ZPhy..235..339T} (For a review see \citet{2002LNP...602....1D}). 

To resolve this ``paradox'', we consider clustering for increasing $b$. As $b$ increases, galaxies become increasingly clustered and large voids are created. These voids have the effect of insulating clusters from the energy and galaxy reservoir that is the rest of the universe. The net effect is that the universe breaks up into a collection of galaxy clusters. Each cluster is a microcanonical ensemble that does not exchange significant energy or galaxies with the rest of the universe.

The measured value of $b$ for the local universe of redshifts $z \lesssim 0.1$ is about 0.86 for large $8^{\circ}$ angular cells \citep{2005ApJ...626..795S} which indicates that $b \approx b_{crit}$ and suggests that clustering is fairly advanced in the local universe. The average matter density(including dark matter) of the universe is $3.5\times 10^{10}m_{\odot}$Mpc $^{-3}$ \citep{2007ApJS..170..377S}. For comparison, the mass and radius of a typical rich cluster are about $10^{15} m_{\odot}$ and 2 Mpc, making it approximately 1000 times denser than the average universe. To find a lower bound to the inter-cluster spacing, consider a simple model in which rich clusters are each surrounded by a void where no galaxies exist. Then each void would be approximately a sphere that would contain roughly $10^{15}m_{\odot}$ if the matter were initially distributed uniformly throughout the universe. The radius of such a sphere is approximately 19 Mpc, which suggests that the average massive cluster separation is at least 19 Mpc.

For an average galaxy with a peculiar velocity of about 1000 km/s to move from one cluster to another over a distance of 19 Mpc would require about 19 Gyr, which is longer than the age of the universe. The long transit times involved suggest that such clusters are no longer likely to exchange galaxies. At separations of 19 Mpc, the interaction potential energy of a pair of galaxies of $\sim 10^{12} m_\odot$ each is approximately 4 orders of magnitude lower than the kinetic energy of a single galaxy with a peculiar velocity of 1000 km/s which indicates that at large values of $b$, clusters would exchange energy more effectively by exchanging galaxies rather than by long range gravitational interactions. But galaxy exchange now takes too long. Hence at $b \gtrsim b_{crit}$, clusters are approximately microcanonical ensembles that are likely to be virialized\citep{2004ApJ...608..636L} and have a negative heat capacity. When most galaxies are bound to a cluster in the case of large $b$, the total heat capacity is negative, and hence the specific heat per galaxy is negative.

To see this, we can use the multiplicity function derived from equation (\ref{1-eq2}) which gives the probability that a physical cluster contains $N$ galaxies (see \citealt{2000dggc.conf.....S}, equation 28.97):
\begin{equation} \label{3-eq3}
\eta(N,b) = \frac{(N b)^{N-1}}{N!}e^{-Nb} \mathrm{\ for\ } N=1,2,3,\ldots
\end{equation}
which is a truncated Borel distribution. From equation(\ref{3-eq1}), the total heat capacity of these clusters is therefore
\begin{equation} \label{3-eq3a}
\frac{3}{2}\left(1+4b-6b^{2}\right)\sum_{N=1}^{\infty}\overline{N}N_{0}\eta(N,b) = \frac{3}{2}\overline{N}N_{0}\left(1+4b-6b^{2}\right)
\end{equation}
where $N_{0}$ is the total number of clusters having an average number $\overline{N}$ of galaxies per cluster. The multiplicity function sums to unity and we can divide the total heat capacity by $\overline{N}N_{0}$ to obtain the average specific heat of the collection of clusters, each of which is a microcanonical ensemble. This gives the same result as equation (\ref{3-eq1}), so the effective specific heat of this ensemble also becomes negative for $b > b_{crit}$ in equation (\ref{3-eq2}).

When $b \to 1$, $f(N) \to 0$ for $N > 0$. The void probability goes to 1 and all galaxies will be bound to a single cluster that can be represented as an isothermal sphere\citep{2003MNRAS.345..552B}. In that limit, the energy and galaxy reservoir is effectively depleted, and the entire system effectively becomes a single microcanonical ensemble.

This transition, as $b$ increases, from a single grand canonical ensemble where the average cluster has a positive specific heat per galaxy through a collection of microcanonical ensembles with negative specific heat per galaxy to a single virialized microcanonical ensemble is an effect of increasing gravitational inhomogeneity. It is rather remarkable that equation (\ref{3-eq1}) can describe this entire ensemble transition in a smooth manner. Perhaps the reason is that properties of the system for $b > b_{crit}$ are essentially determined by its properties for $b < b_{crit}$.

\subsection{Robustness to Mergers}

Galaxies are known to interact and merge with each other. Over time, we can expect the average number of galaxies in a given comoving volume to decrease due to mergers, and hence galaxy mergers will change the distribution of galaxies. We can classify galaxy mergers by the masses of the progenitors $m_1$ and $m_2$. Minor mergers are mergers where one galaxy is much less massive than another galaxy such that $m_1 \ll m_2$ or $m_2 \ll m_1$. In such mergers, the resulting galaxy will have a position that is close to the centre of mass of the larger of the two galaxies and the only change to the distribution is the reduction of the number of low mass galaxies. The other class of mergers are major mergers where both progenitors are of comparable mass such that $m_1 \approx m_2$, and the position of the resulting galaxy is approximately at the midpoint between the two. Major mergers will change the spatial distribution of galaxies more drastically. The average number density $\overline{N}$ decreases with time and the average mass of each galaxy increases with time. This affects the rate at which $b$ changes.

To follow the change in the distribution of galaxies due to mergers, we consider the distribution of nearest neighbour galaxy pairs whose members are separated by a distance $2\mathbf{r}$. We denote the separation between the midpoints of two such pairs by $\mathbf{R}$. Hence the interaction between two galaxy pairs, each with separations $2\mathbf{r}_{1}$ and $2\mathbf{r}_{2}$, is given by
\begin{eqnarray} \label{3-eq6}
\phi(\mathbf{r}) = -\frac{G m^2}{|\mathbf{R}|}& &\left(\frac{|\mathbf{R}|}{|2\mathbf{r}_1|}+\frac{|\mathbf{R}|}{|2\mathbf{r}_2|}+\frac{|\mathbf{R}|}{|\mathbf{R}+\mathbf{r}_1+\mathbf{r}_2|}+\frac{|\mathbf{R}|}{|\mathbf{R}-\mathbf{r}_1+\mathbf{r}_2|}\right.\nonumber \\
& &\left. +\frac{|\mathbf{R}|}{|\mathbf{R}+\mathbf{r}_1-\mathbf{r}_2|}+\frac{|\mathbf{R}|}{|\mathbf{R}-\mathbf{r}_1-\mathbf{r}_2|}\right)
\end{eqnarray}
where the term in brackets can be viewed as a modification to the point mass potential. By averaging over possible values of $\mathbf{r}_1$ and $\mathbf{r}_2$, we can use equation (\ref{3-eq6}) to define a modification $\kappa(|\mathbf{R}|,\langle \mathbf{r} \rangle)$ to the potential where $|\mathbf{R}|$ is the separation between two pairs. From section \ref{s2-mod}, The modification due to such a potential will enter into the form of $b_0$ and hence the resulting distribution will also follow the GQED. By assuming that all galaxies will eventually merge, we note that when all galaxies have merged once, the spatial distribution of the resulting galaxies also follows the GQED. Hence the distribution of galaxies is robust to mergers.

\subsection{The Time Evolution of $b$}
Since there is no final equilibrium state for a classical infinite gravitating system, its clustering increases as described by the increase of $b$ with time. From the thermodynamic variables in section \ref{s1-intro}, we can write (cf. \citet{1992ApJ...391..423S})
\begin{equation} \label{3-eq7}
b = b_0(\overline{N}) P T^{-4}.
\end{equation}
Then by taking the expansion of the universe to be an adiabatic process,
\begin{equation} \label{3-eq8}
0=dU+PdV=\frac{3}{2}(1-2b)\left[Td\overline{N}|_{T,P}+\overline{N}dT|_{\overline{N},P}\right]-3\overline{N}Tdb+\overline{N}T(1-b)\frac{dV}{V}.
\end{equation}

If galaxy mergers do not occur, $d\overline{N} = 0$. With the relation $dV/V = 3da/a$, we arrive at a relation relating the scale length, $a$, of the universe with $b$ and $\overline{N}$ such that
\begin{equation} \label{3-eq9}
\frac{1+6b}{8b}\frac{db}{da} = \frac{1-b}{a}.
\end{equation}
Integrating equation (\ref{3-eq9}) we get\citep{1992ApJ...391..423S}
\begin{equation} \label{3-eq10}
\frac{b^{1/8}}{(1-b)^{7/8}} = \frac{a}{a_{*}}
\end{equation}
where $a_{*}$ is a constant of integration given by the initial state.

Galaxies however will merge, and hence we can expect that $d\overline{N} \neq 0$. This gives
\begin{equation} \label{3-eq11}
\frac{1+6b}{8b}\frac{db}{da} = \frac{1-b}{a} + \frac{1-2b}{2\overline{N}} \left.\frac{d\overline{N}}{da}\right|_{T,P}.
\end{equation}
From the definition of $b$ given in equation (\ref{1-eq3}), we note that $b$ depends on $\overline{n}$ and hence $\overline{N}$ and $V$. Denoting $\overline{m}$ as the average mass of a galaxy, by conservation of mass we have $\overline{m}\overline{N} = $ constant for an ensemble of comoving volumes. This gives
\begin{equation} \label{3-eq12}
\frac{db}{b} = -6\zeta_{\star}\left.\frac{d\overline{N}}{\overline{N}}\right|_{T,P}.
\end{equation}

Then from equations (\ref{3-eq11}) and (\ref{3-eq12}), the rate of change of $b$ is given by
\begin{equation} \label{3-eq14}
\left(\frac{1+6b}{8b}+\frac{1-2b}{12b\zeta_{\star}(\epsilon/R_1)}\right)\frac{db}{da} = \frac{1-b}{a}.
\end{equation}
where $\zeta_{\star}$ is a term that is given by
\begin{equation} \label{3-eq13}
\zeta_{\star}(\epsilon/R_1) = 1+\frac{1}{18}\frac{\epsilon}{R_1} \frac{\zeta'(\epsilon/R_1)}{\zeta(\epsilon/R_1)}
\end{equation}
for the case of the isothermal halo with $\zeta(\epsilon/R_1)$ given by equation (\ref{2-eq2a}). Here $\zeta'(\epsilon/R_1)$ is the derivative of $\zeta(\epsilon/R_1)$ with respect to $\epsilon/R_1$. In this case $\zeta_{\star}$ is a factor of order unity that varies between $1$ and $17/18$.

\begin{figure}[tbp]
\begin{center}
\includegraphics[width=\textwidth]{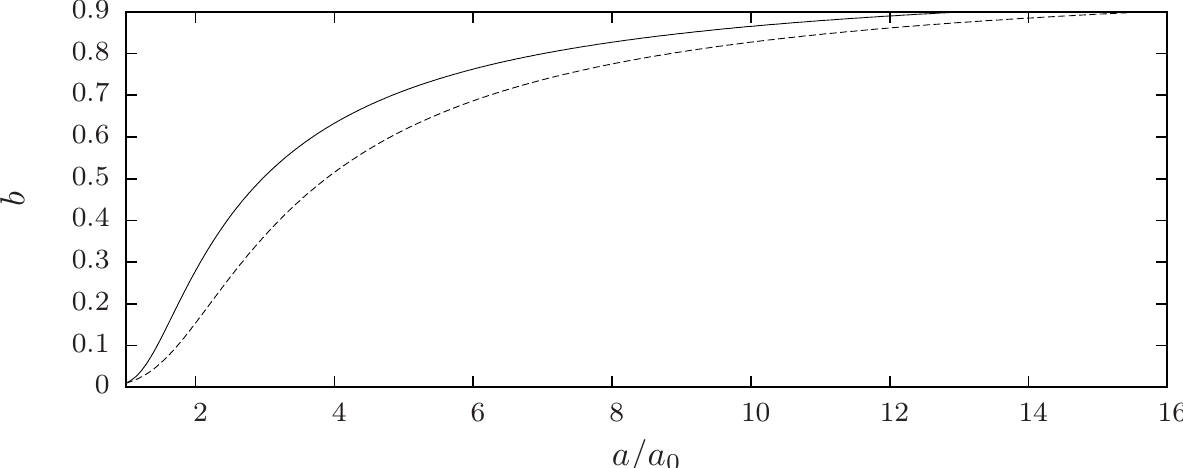}
\caption{The evolution of $b$ with respect to scale length for an initial value of $b=0.01$ at $a/a_{0}=1$. The solid line shows the case without mergers(equation \ref{3-eq10}) and the dashed line shows the case with mergers(equation \ref{3-eq15}).}
\label{3-fig2}
\end{center}
\end{figure}

In the limit of a small halo, $\epsilon/R_1 \to 0$ and $\zeta_{\star}(\epsilon/R_1)\to 1$ and equation (\ref{3-eq14}) integrates to become
\begin{equation} \label{3-eq15}
\frac{b^{5/24}}{(1-b)^{19/24}} = \frac{a}{a_{*}}.
\end{equation}
In both the non-merging and merging cases, we note that $b$ increases as $a$ increases with $b$ increasing faster in the case with no mergers than in the case with mergers. We compare the two cases in figure \ref{3-fig2}.

\section{Simulations and Observations}
The results of the GQED have been compared with $N$-body simulations and various observations, and there is very good agreement between predictions, simulations and observations without having to introduce additional(e.g. non-gravitational) parameters.

\subsection{$N$-body Simulations}
\begin{figure}[tbp]
\begin{center}
\includegraphics[width=\textwidth]{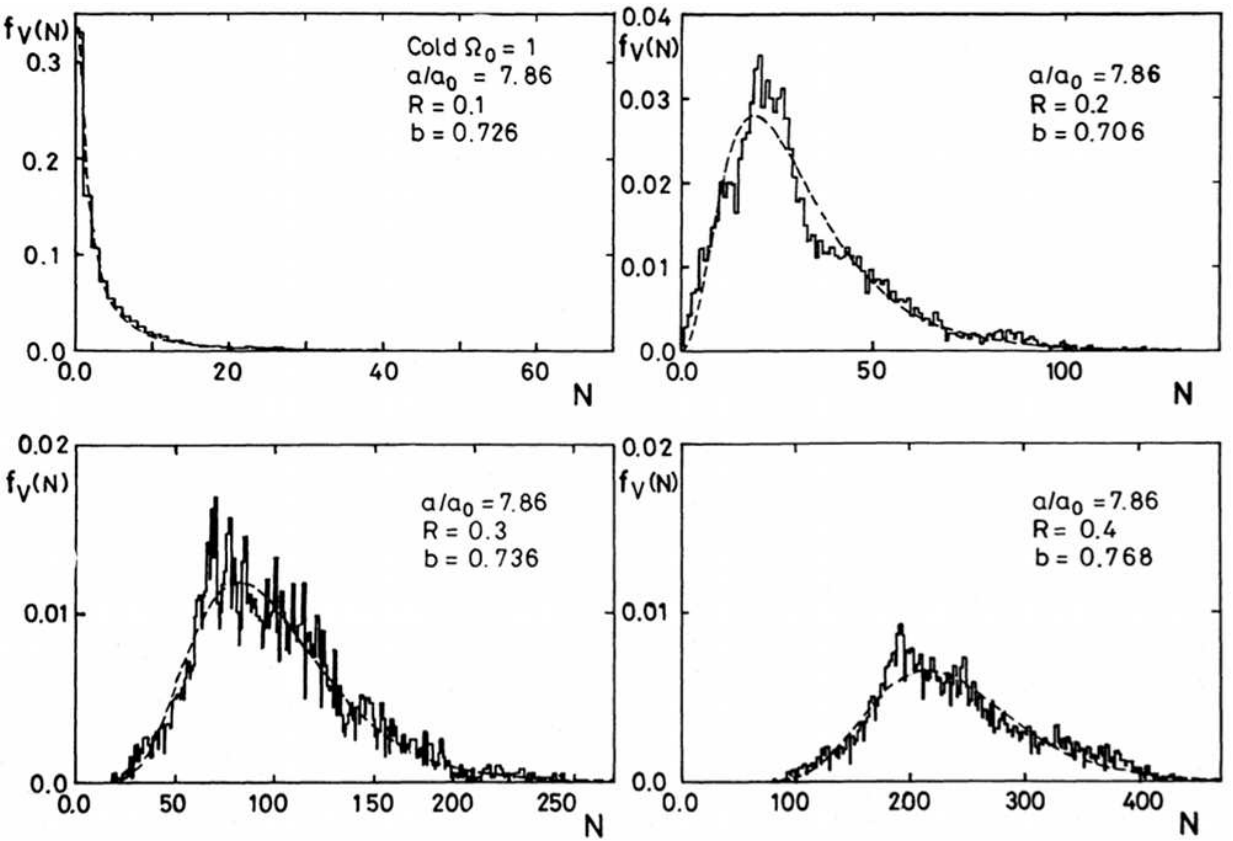}
\caption{$f_{V}(N)$ for a single simulation of $\Omega_{0} = 1$ and cold initial conditions for the expansion factor $a/a_{0}=7.86$ from \protect\citet{1988ApJ...331...45I}. Various cell radii $R$ between 0.1 to 0.4 are plotted. The solid lines are from the simulation and the dashed lines are from equation (\ref{1-eq2}).}
\label{4-fig1}
\end{center}
\end{figure}

\begin{figure}[tbp]
\begin{center}
\includegraphics[width=\textwidth]{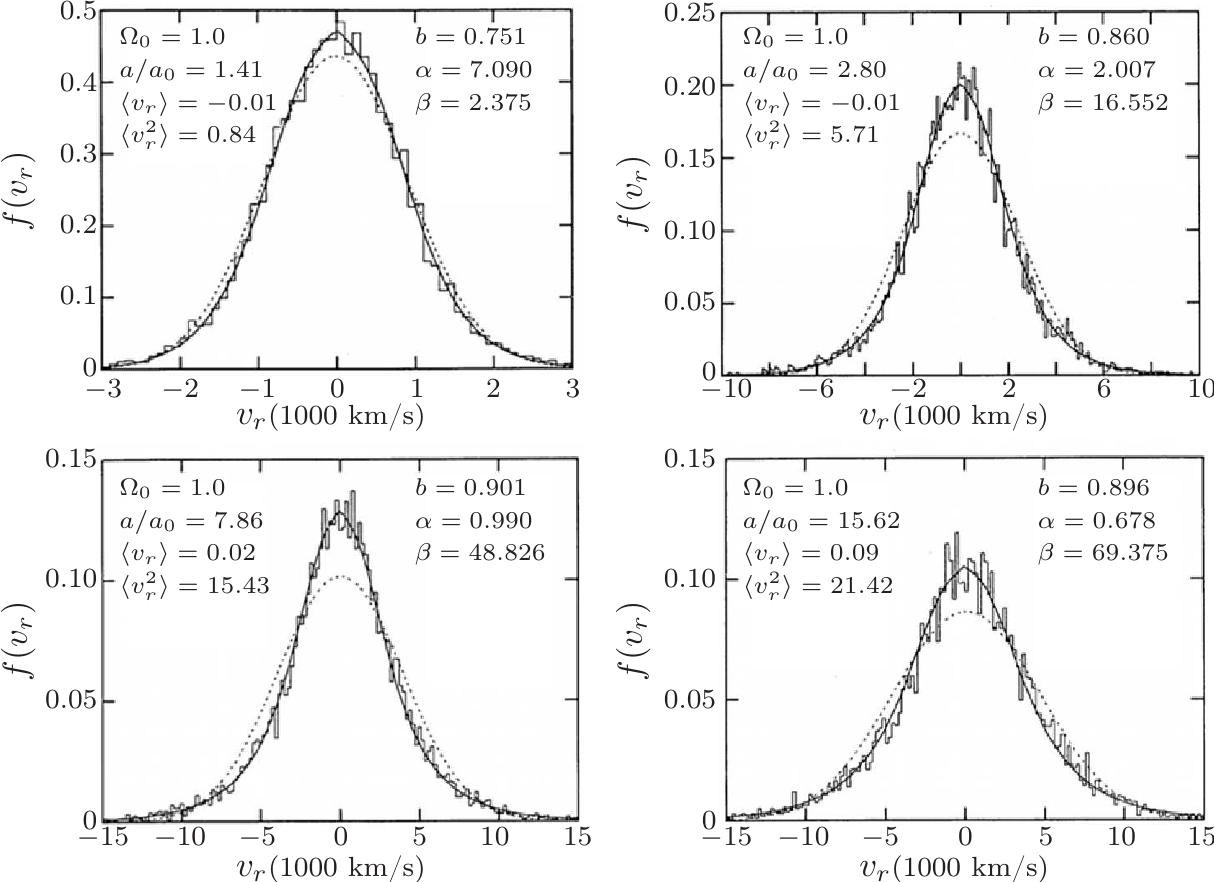}
\caption{The radial velocity distribution $f(v_{r})$ for a single simulation of $\Omega_{0} = 1.0$ from \protect\citet{1992ApJ...386....9I}. The histogram is data from the simulation. Various expansion factors $a/a_{0}$ are plotted. The solid lines are from equation (\ref{1-eq13}) and the dashed lines are Maxwell-Boltzmann distributions with the same $\langle v_{r}^{2} \rangle$.}
\label{4-fig2}
\end{center}
\end{figure}

A series of computer simulations \citep{1988ApJ...331...45I,1990ApJ...356..315I,1992ApJ...386....9I,1993ApJ...403..476I} examined the various aspects of the GQED in a comoving volume with $N$ between 4000 and 10000. The spatial distribution and peculiar velocity distributions were examined for cases where galaxies had the same mass and for various mass spectra. The primary results of these simulations were that the distributions of galaxies follow the GQED very well for identical masses, and even better for a more realistic case where galaxies have different masses. Figure \ref{4-fig1} shows a typical example for identical masses and figure \ref{4-fig2} shows examples of velocity distributions.

\subsection{Observations of $f_{V}(N)$}
\begin{figure}[tbp]
\begin{center}
\includegraphics[width=\textwidth]{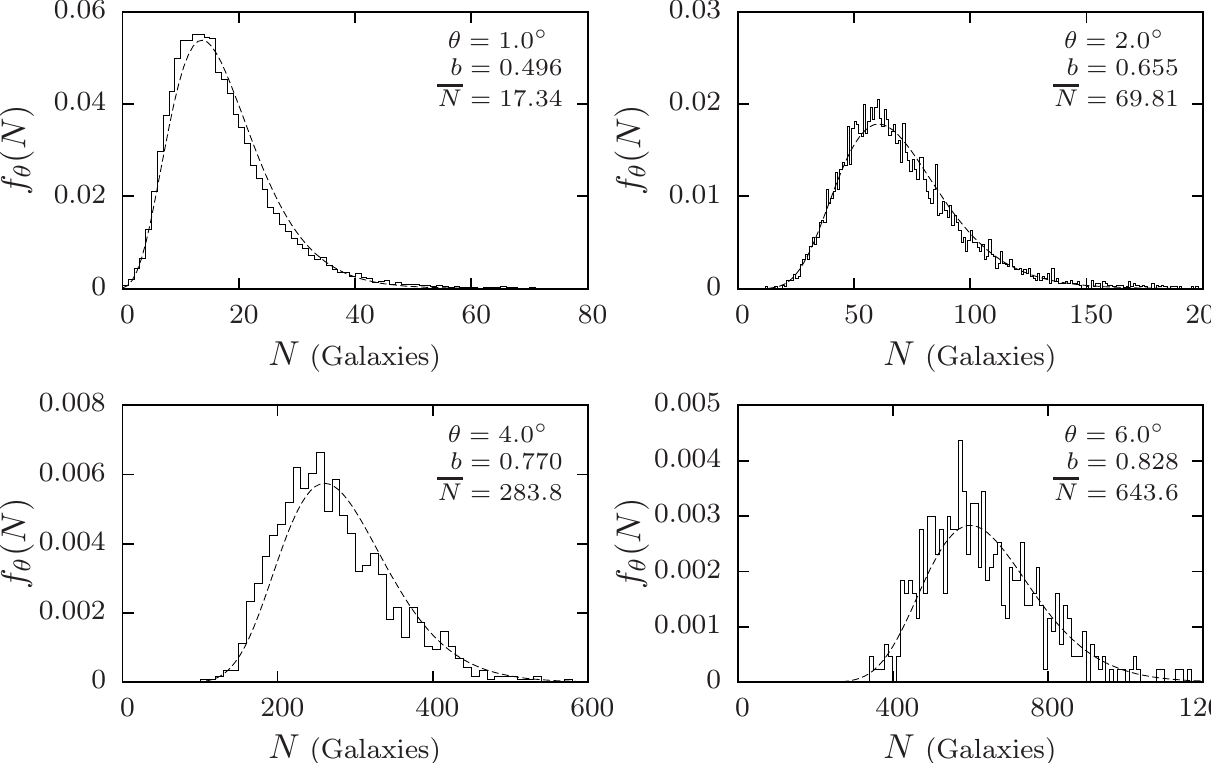}
\caption{The angular projected GQED $f_{\theta}(N)$ for square cells with different values of $\theta$ from \protect\citet{2005ApJ...626..795S}. The solid histogram is data from observations. The dashed curve is from equation (\ref{1-eq2}). Values of $b$ and $\overline{N}$ were found directly from the observations.}
\label{4-fig3}
\end{center}
\end{figure}

There have been many observations of $f_{V}(N)$ using various galaxy catalogs. Some recent ones(which also give earlier references) include
\begin{enumerate}
\item An analysis of 2-dimensional angular cells \citep{2005ApJ...626..795S}, figure \ref{4-fig3}, using data from the 2MASS survey, an all-sky survey in the infrared with up to 439754 galaxies. A good fit to predictions was found with $b$ approaching $b_{crit}$ at large cell sizes.

\item A 3-dimensional analysis of the Pisces-Perseus supercluster with 4501 galaxies\citep{1998ApJ...509..595S} found $\overline{N} = 3.38$ and $b = 0.80$ for cells of $10 h^{-1}$ Mpc where $h$ is the reduced Hubble constant such that $h = H_{0}/100$. The $\chi^{2}$ value was found to be 0.089, indicating a relatively good fit to the GQED.

\item At high redshifts, \citet{RST2009} found that the projected spatial distribution follows the GQED out to a redshift of $z=1.5$ using the GOODS survey catalog. Due to the small sample size and limited sky coverage, large differences between the North and South fields in the GOODS catalog are evident. Nevertheless, there is a good agreement between the form of the GQED and the observed spatial distribution of galaxies even at these high redshifts.

\end{enumerate}

\subsection{Observations of $f(v)$}

\begin{figure}[tbp]
\begin{center}
\includegraphics[width=\textwidth]{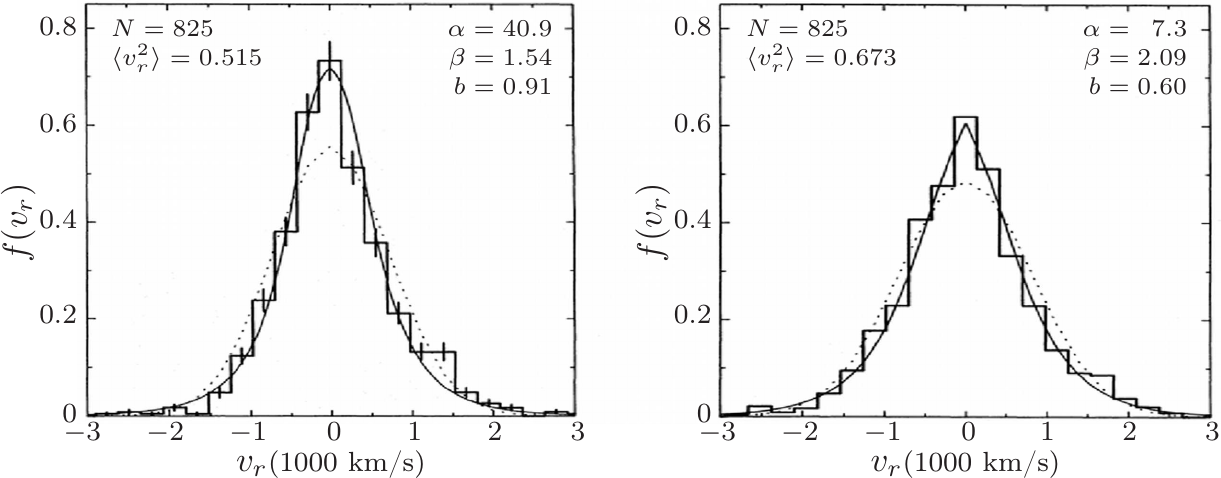}
\caption{The observed radial velocity distribution $f(v_{r})$ from \protect\citet{1996ApJ...461..514R}. The histogram is data from observations. The solid lines are from equation (\ref{1-eq13}) and the dotted lines are best fit Maxwell-Boltzmann distributions. The left panel shows the radial velocity distribution corrected for a bulk motion of 599 km/s while the right panel shows the radial velocity distribution corrected for an extra Hubble expansion with an effective local Hubble constant of $H_{0} = 92$ km/s/Mpc.}
\label{4-fig4}
\end{center}
\end{figure}

The radial velocity distribution was analysed observationally \citep{1996ApJ...461..514R} using the Matthewson catalog of spiral galaxies. Since radial velocities of galaxies require a secondary distance indicator and are not easily obtained, the catalog is relatively small with 1353 galaxies. Despite the small sample, a relatively good fit to the radial velocity distribution of equation (\ref{1-eq13}) was found as illustrated in figure \ref{4-fig4}.

\section{Conclusion}
Although the cosmological many-body problem is essentially a long range gravitating system, the expansion of the universe cancels the long-range gravitational mean field so that interactions effectively have a finite range. Using this property and its consequences, we have described the spatial and velocity distributions of particles in such a system. Its fundamental particles are galaxies. In the simplest case the spatial distribution of galaxies can be characterised by just the average number of galaxies in a cell, $\overline{N}$, and a clustering parameter $b$, which is essentially the ratio of interaction and kinetic energies. This assumes that galaxies are equal point masses, and that pairwise interactions are the dominant form of interactions. Relaxing these assumptions introduces minor higher order corrections into $b$. However the form of the spatial distribution does not change, suggesting it is very robust to a wide range of physical conditions.

The velocity distribution function can be derived consistently from the spatial distribution which can also be used to obtain the volume integrals of the $N$-point correlation functions.  From the thermodynamic quantities the time evolution of $b$ can be derived, describing the rate at which clusters form. Towards increasing $b$, the grand canonical ensemble gradually breaks up into a collection of microcanonical ensembles, each of which are virialized clusters that individually have negative specific heat such that the average specific heat goes negative when $b$ is greater than a critical value $b_{crit} = 0.8604$.

Finally, comparisons with $N$-body simulations and observations show that the GQED does indeed describe the physical world, at both low and high redshifts and for both spatial and velocity distributions.

\textbf{Acknowledgments}: We are both grateful to the organisers of
the Les Houches Summer School for an environment of stimulating
discussion. We particularly thank Thierry Dauxois and Michael
Kastner for their comments on negative specific heat.


\end{document}